# Dynamics of liquid crystalline domains in magnetic field

Giorgia Tordini, P. C. M. Christianen, J.C. Maan
NSRIM, High Field Magnet Laboratory, University of Nijmegen, Toernooiveld 7, 6525 ED Nijmegen, The Netherlands
Tel: (+31) (0) 24 3652087 Fax: (+31) (0)24 3652440 E-mail: giorgia@sci.kun.nl

## Abstract

We study microscopic single domains nucleating and growing within the coexistence region of the Isotropic (I) and Nematic (N) phases in magnetic field. By rapidly switching on the magnetic field the time needed to align the nuclei of sufficiently large size is measured, and is found to decrease with the square of the magnetic field. When the field is removed the disordering time is observed to last on a longer time scale. The growth rate of the nematic domains at constant temperature within the coexistence region is found to increase when a magnetic field is applied.

**Keywords**: Liquid crystals, phase transition, dynamics, alignment, magnetic field.

## 1 Introduction

The study of the properties of liquid crystals (LCs) has fascinated scientists since the early sixties [1] and has lead to new theories and new applications, making that liquid crystals have entered the daily life. Recently we have found that polymer liquid crystals can be aligned in a magnetic field, but only when the temperature is swept through the I-N phase transition at fixed field, while no effect is observed at fixed temperature upon sweeping the field [2]. Clearly the dynamical behaviour in a magnetic field around the phase transition is responsible for this observation; thus we study this region in more detail, using a simple LC as a model system. This phase transition region has attracted great attention both from the theoretical and the experimental point of view [3-7].

Upon cooling a nematic LC from the isotropic phase to the LC phase the critical temperatures are: the temperature at which the nematic phase changes from unstable to metastable, $T_1$, the temperature at which the nematic phase and isotropic phase are equally stable, $T_2=T_{IN}$, and the temperature at which the isotropic phase becomes completely unstable, $T_3$. Nematic nuclei are formed around $T_{IN}$ and rapidly grow in size and number in an isotropic background, within the coexistence region, until they coalesce into the fully developed nematic phase [8,9]. We measure with polarized microscopy the behaviour of the nematic domains in this transition region, both at constant magnetic field (growth rate) as well as in transient magnetic fields (dynamical magnetic alignment). We show that each nematic domain orients parallel to the field direction via a rotation of the domain as a whole, with a time $\tau_{al}$ depending on the field strength, as expected. Smaller domains need a higher field to align than bigger ones. Upon switching off the field the domains gradually disorient by thermal fluctuations in a time $\tau_{dis}$ that is much longer than $\tau_{al}$. The disordering time increases upon increasing domain dimensions.

At fixed temperature, after a rapid quench, domains grow with time according to a power law, with exponent around 0.5 at zero field, as theoretically predicted [10, 11]: upon increasing the steady magnetic field the exponent increases (faster growth) and it tends to 1 at 2T.

## 2 *Experimental details*

A LC in a cuvette is mounted in a thermal oven placed in an electromagnet (max field 2T). The sample is illuminated with a He-Ne laser (543.5nm) and the light transmitted between crossed polarizers is monitored with a video camera. The oven allows forced and regulated heating and cooling, giving high temperature stability ($\Delta T \sim 0.001$ °C) and a rapid rate of change (2K/m) for small quenches (0.1-0.3K). A 10X magnification microscope objective is connected to the oven, focusing the light from the sample on a CCD camera. The two polarizers at 90 degrees before and after the sample block all the non birefringent light, allowing to measure the alignment of the nematic domains by studying their intensity. By orienting the polarizers at 45 degrees with respect to the field direction fully aligned domains appear as the brightest. The pure liquid crystal material, MLC6610 from *Merck*, is filled in the isotropic phase by capillarity in a glass cell of diameter ~ 500µm; this thickness is chosen to reduce surface anchoring effects. With this experimental configuration we can acquire with micrometer resolution images and movies, which we process for each pixel in each frame quantitatively.

## 3 *Theoretical background*

Liquid crystalline domains containing N molecules exhibit an anisotropic diamagnetic susceptibility $N\Delta\chi$ leading to an extra energy dependent on their orientation. Order will therefore be induced if the magnetic energy is larger than the thermal energy, i.e.

$$N \frac{\Delta \chi B^2 \cos^2(\theta)}{\mu_0} \geq kT \qquad (1)$$

The typical domain size for which this condition is fulfilled is for most LC of the order of a micrometer in a few Tesla.

By rapidly switching on the field in the coexistence region, we study the velocity with which the nematic domains align, and their disordering when the field is switched off.

## 4 *Results and discussion*

Upon slowly cooling from the isotropic phase, the system enters in the metastable [1] coexistence region, where clusters of nematic phase nucleate and start growing. In this region we can stabilize nematic nuclei of a certain size for several minutes, by keeping T constant. Under these conditions, when the magnetic field is switched on in less than a second to a certain value (between 0.1 and 2 Tesla), we observe the domain rotation towards the direction of the applied field.

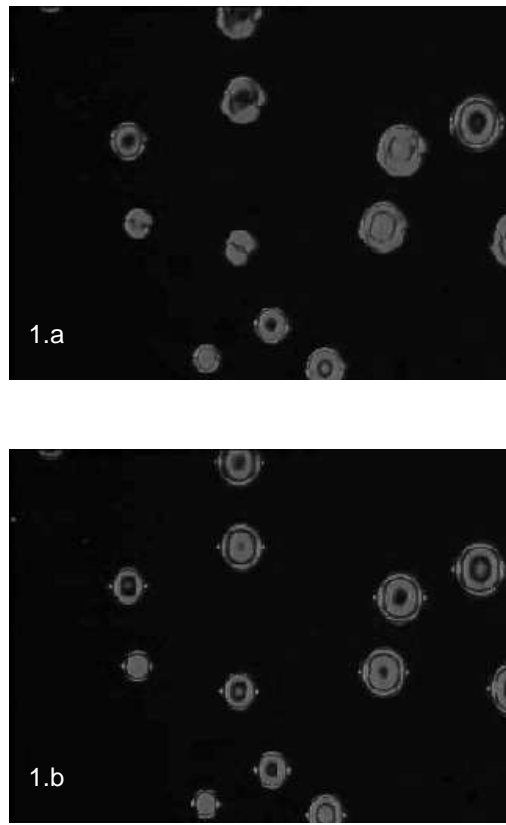

Fig. 1. Nematic domains aligning in magnetic field. 1.a: at a field intensity of 0.1 Tesla only domains bigger than 10 µm align. 1.b: at a field intensity of 0.2 Tesla also smaller domains (~5 µm) align.

The rotation takes place in few seconds, but at low fields it is slower than the variation of the field. Fig. 1a and 1b shows that, at different values of the applied magnetic field, the bigger domains start to orient first, followed by the smaller ones at higher fields. We concentrate in the following on domains that are relatively isolated from each other and initially oriented in the plane of the magnetic field. In Fig. 2a we show the change of orientation for these individual domains as a function of time for different magnetic fields. When a domain has an initial orientation at an angle, in the plane of the field, almost perpendicular to field direction, it can either rotate left of right towards the field direction. Since for small deviations from 90° the gain in energy is very small, the domain may fluctuate around its initial orientation for a while, until, by thermal fluctuation, it moves sufficiently away from this unstable equilibrium. At this point a domain orients towards the field direction with a velocity that rapidly increases as the field increases. Therefore all curves in Fig. 2 show some initial 'waiting time', after which they start rotating. We determine the relaxation time $\tau_{al}$ only from the rapidly decaying part of the curves. We find that $\tau_{al}$ decreases very rapidly with increasing field and above 0.2 T it is already shorter than our time resolution.

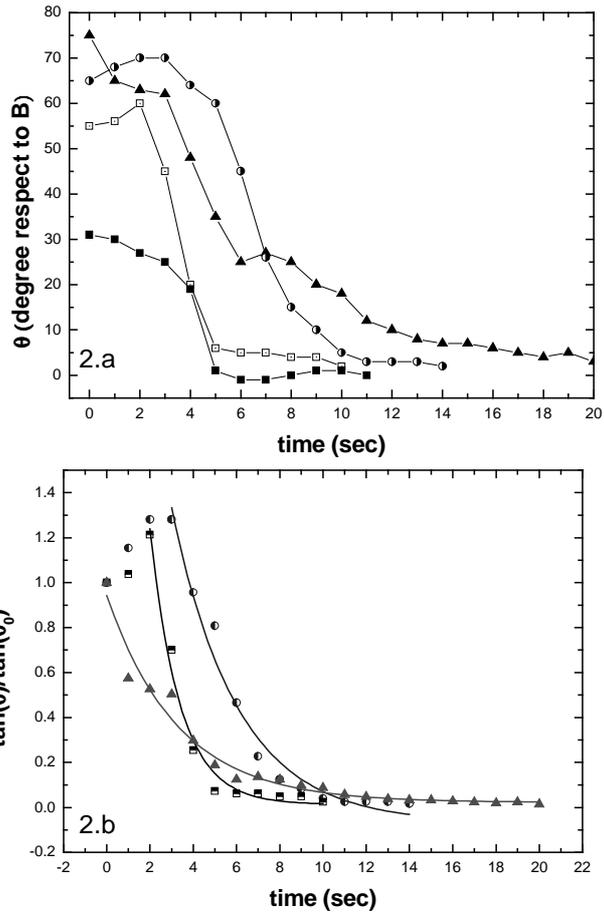

Fig. 2. Rotation of nematic domains to align in the field direction ($\theta=0$). 2a: open squares: alignment in 2 Tesla for a domain of initial size $L(t_0)=11.6\mu m$; full squares: alignment in 2 Tesla, $L(t_0)=10.5\mu m$; circles: alignment in 0.2 Tesla, $L(t_0)=12\mu m$; triangles: alignment in 0.1 Tesla, $L(t_0)=12.9\mu m$. 2b: $\tan(\theta)/\tan(\theta_0)$ versus time. From the exponential fit we extract $\tau_{al}$.

To be more quantitative we write the equation of motion for magnetic orientation, which is the balance of the magnetic torque and the hydrodynamic torque [13]:

$$Q\frac{d\theta}{dt} = -\frac{1}{2}V\Delta\chi\mu_0 B^2 \sin(2\theta) \qquad (2)$$

where V is the domain volume and $Q$ is the friction: $Q = 6\pi\eta R^3$.

A solution of the equation of motion, assuming perfectly spherical domains, is

$$\tan(\theta) = \tan(\theta_0)\exp\left(\frac{-t}{\tau_{al}}\right) \text{, with } \tau_{al} = \frac{6\eta}{\mu_0\Delta\chi B^2} \qquad (3)$$

where $\eta$ is the rotational viscosity of the material.

In Fig. 2b we plot $\tan(\theta)/\tan(\theta_0)$ and determine $\tau_{al}$ from the rapidly decaying part. $\tau_{al}$ is of the order of 4 seconds for 0.1 T and rapidly decreases by increasing the magnetic field, as expected from eq. (3). Within our experimental accuracy we see no size dependence of $\tau_{al}$, as predicted by eq. (3). The reason for this independence on size is that both the driving and the friction torque are proportional to the volume of the domain.

Switching off the magnetic field instantaneously after the domains have been aligned, we study the relaxation of the domain as a function of time. In this case there is no driven motion and only thermal fluctuations drive the domain to a random orientation. This statistical process is found to be faster for small domains than for big domains, which is reasonable since, for increasing domain size, the thermal disorienting force on the molecules at the interface becomes smaller compared to the number of oriented molecules inside the cluster. We measure a disorientation time of ~3 minutes for domains of 13μm (Fig. 3a and 3b), 7 minutes for domains of ~25μm size while domains of size ~50μm remain aligned up to 15 minutes. $\tau_{dis}$ is thus roughly found to be proportional to L.

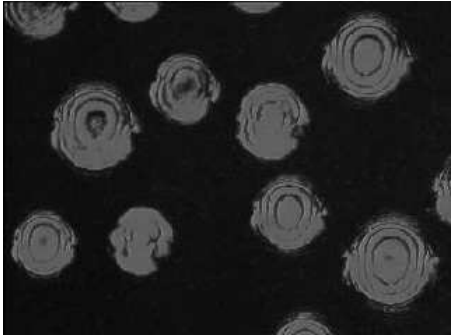
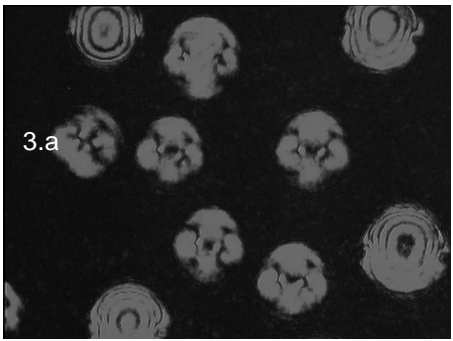

Fig. 3. Disordering of nematic domains after being aligned in magnetic field. 3.a: domains after 180 sec: small domains (~10μm) have already disordering. 3b: after 420 sec only domains of size ~ 50μm are still aligned.

We also studied the growth velocity of domains by under-cooling rapidly from the isotropic phase (0.1K/min) to a temperature slightly below the $T_{I-N}$, and then stabilize the temperature. Although in this condition the probability of nucleation is not high, after some minutes we can see formation of clusters of nematic phase and their subsequent growth. In order to control the time at which the phase transition starts we perform slow and small quenches into a temperature region between $T_1$ and $T_3$ with a quench depth is between 0.1°C and 0.3°C and wait ~100 seconds in order to reach a constant temperature to study the growth of the nematic domains. We do not see any difference in the growth behaviour in the small quench and no quench experiments.

We note the presence of two different regimes during the growth: an incubation phase at the beginning followed by and an expansion (growth) phase (Fig.4). We have fitted the growth phase with a power law

$$L(t) \sim (t-t_0)^\alpha \qquad (4)$$

[10,11] with $\alpha$ a fitting parameter. In zero field we find $\alpha$ to be 0.5 ± 0.2, as theoretically predicted. In magnetic field $\alpha$ tends to increase from 0.5 to 1 between 0.1 Tesla to 2 Tesla (Fig. 5). Although there is a large error on the fitted exponent, mainly caused by the uncertainty in $t_0$, i.e. where the incubation phase finishes, the trend of increasing growth velocity in magnetic field is clear.

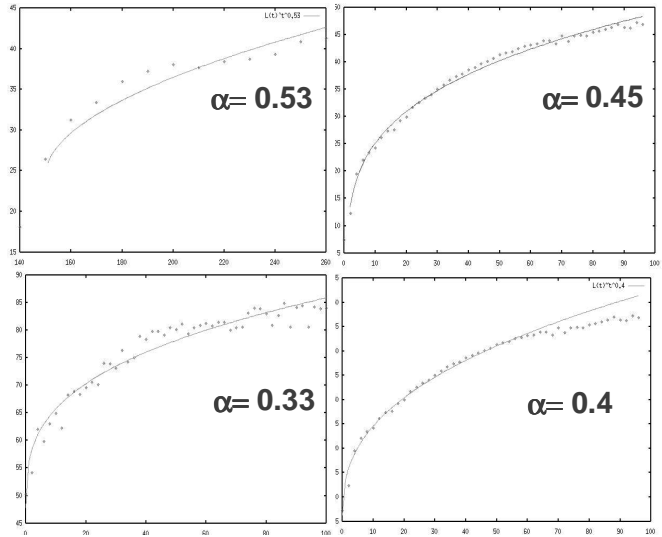

Fig. 4. Size of a growing domain (a.u.) in time (sec) in no magnetic field. Note the presence of two regimes in the growth: incubation and expansion. The line is the fit with the power law: $L(t) \sim (t-t_0)^\alpha$ where $\alpha$ is found 0.5±0.2.

This is an unexpected result, for which at present we have no satisfactory explanation. Since the magnetic energies are very small, we expect this faster growth to be caused by kinematical effects, rather than by a change in thermodynamic equilibrium.

# 5 *Summary*

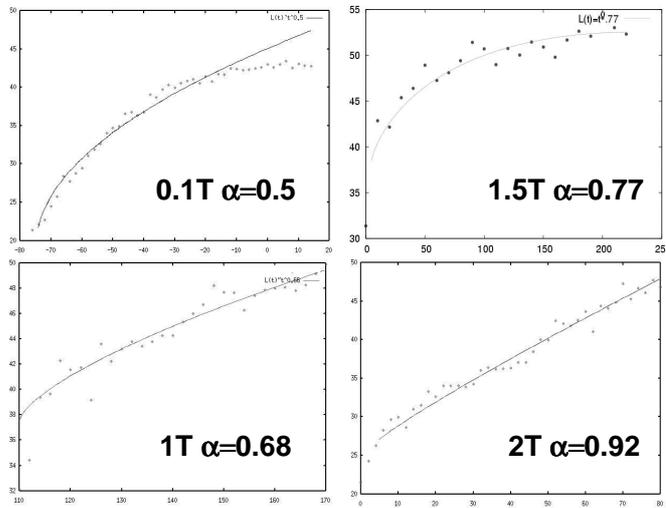

Fig. 5. Size of a growing domain (a.u.) in time (sec) in magnetic field. The expansion phase is fitted by a power law $L(t) \sim (t-t_0)^\alpha$ with $\alpha$ increasing by increasing the field.

We have studied for the first time the dynamics of single nematic domains in magnetic field, near the Isotropic-Nematic phase transition. We are able to monitor the growth in its early stage, which we identify as an incubation state, and which is followed by stable growth. In this expansion phase, a $t^{1/2}$ growth law is observed as theoretically predicted. In magnetic field a faster growth described with a higher exponent is observed, which is not yet theoretically understood.

We have measured the time of alignment in magnetic field at constant temperature for domains of different size, verifying the predicted behaviour: $\tau_{al} \propto 1/B^2$.